# Enhancing Crystal Structure Prediction by decomposition methods based on graph theory

Hao Gao,[1] Junjie Wang,[1] Yu Han[1] and Jian Sun[1,*]

*1 National Laboratory of Solid State Microstructures, School of Physics and Collaborative Innovation Center of Advanced Microstructures, Nanjing University, Nanjing 210093, China*

**ABSTRACT**

Crystal structure prediction algorithms have become powerful tools for materials discovery in recent years, however, they are usually limited to relatively small systems. The main challenge is that the number of local minima grows exponentially with system size. In this work, we proposed two crossover-mutation schemes based on graph theory to accelerate the evolutionary structure searching. These schemes can detect molecules or clusters inside periodic networks using quotient graphs for crystals and the decomposition can dramatically reduce the searching space. Sufficient examples for test, including the high pressure phases of methane, ammonia, $MgAl_2O_4$ and boron, show that these new evolution schemes can obviously improve the success rate and searching efficiency compared with the standard method in both isolated and extended systems.

*corresponding author. Email: jiansun@nju.edu.cn

**INTRODUCTION**

In the past ten years, crystal structure prediction has been widely applied to many fields in physics, chemistry and materials science. Many practical structure prediction methods, including random structure searching[1], evolutionary algorithm[2–6] and particle swarm optimization[7], basin hopping[8], minima hopping[9], simulated annealing[10], metadynamics[11], Bayesian optimization[12–14] etc., are designed to search for optimal structures with the lowest energy or enthalpy under target pressures. Combined with Density functional theory (DFT), these methods can not only identify unclear experimental results but also predict structures to drive materials discovery[15,16]. A lot of these predictions have been confirmed by the following experiments[17–20].

Although these structure searching methods have been successful in exploring new materials, most of their applications still focus on relatively small systems because searching for large systems is demanding. The results of crystal structure prediction are mainly influenced by two parts: sampling algorithms for exploring the structural space and local optimization algorithms for relaxing structures. Thus the difficulties in predicting large systems based on first-principles calculations derive mainly from two reasons: one is the exponentially growing complexity of global structure determination and the other is the high time cost of structure relaxations base on DFT. For the second problem, a popular approach is combining machine learning (ML) algorithms with structure prediction methods[12,13,21–26]. These methods train an on-the-fly ML model during prediction processes and use the model to select and relax candidate structures as a reliable surrogate for DFT to accelerate global searches. Compared with the difficulty induced by first-principles calculations, the first problem of the exploration on configuration space is more general and difficult to solve. Deringer *et al.*[25] have employed an advanced machine learning potential to accelerate the searches for allotropes of phosphorus. Their method is able to find black phosphorus which proves that the surrogate potential may be reliable, but fails in finding fibrous phosphorus with a complex crystal structure, mainly caused by the high dimension of the search space.

On the other hand, graph theory has been applied to crystallography long ago[27–29], but its applications in condensed matter physics and material science are still inadequate. Recently, Shi *et al.*[30] proposed a random sampling method with a combination of space group and graph theory. They found a series of complex $sp^3$ carbon polymorphs with large band gaps[31]. Bushlanov *et al.* developed a topological random structure generator[32] to produce structures as seeds of evolutionary structure searching. This generator is based on topological databases containing idealized periodic nets from known crystal structures[29,33] and has remarkable performance improvements. Ahnert *et al.*[34] used modularity optimization technique to decompose atomic networks into modules. Combined with random structure searching, this approach is able to find low-energy hypothetical allotropes of boron and phosphorus[25,34]. These previous works show that graph theory can effectively reduce the complexity of searching space.

In the present paper, we combine graph theory and evolutionary algorithm to enhance crystal structure predictions. Two evolution schemes using structure decomposition is proposed. The first one can automatically identify molecules in crystals and search for molecular crystals efficiently without setting molecular geometry in advance. The other is based on the first scheme, but it uses a community detection algorithm to further decompose extended atomic networks. Extensive tests show that these crossover-mutation schemes have significant advantages compared with the standard scheme in different systems.

## RESULTS AND DISCUSSION

### Quotient graph theory and dimensionality identification

At first, we simply introduce the definition of quotient graph (QG)[27], which is the basis of the structure decomposition methods in this work. QG is a simplified approach to representing crystals including information of atoms and bonds. For a crystal containing $N_{at}$ atoms in the unit cell, the corresponding QG is a labeled and directed graph with $N_{at}$ vertices $\{n_1, n_2, ..., n_{N_{at}}\}$. To distinguish equivalent atoms in different cells, a notation $n_i(\boldsymbol{v})$ is used for representing the atom at $(\boldsymbol{x}_i + \boldsymbol{v})\boldsymbol{h}$, where $\boldsymbol{x}_i$ is the fractional coordination of the $i$th atoms, $h$ is the cell and $v$ is the coordination of the cell. If there is a bond between a pair of atoms $n_i(v')$ and $n_j(v'')$, the QG has an edge $n_i \xrightarrow{v=v''-v'} n_j$ with a label vector $v$. All the equivalent bonds between $n_i(v_0 + v'')$ and $n_j(v_0 + v')$ can be represented by the edge $n_i \xrightarrow{v=v''-v'} n_j$, where $v_0$ is an arbitrary integer vector. It is easy to find that the edge $n_i \xrightarrow{v} n_j$ is equivalent to $n_j \xrightarrow{-v} n_i$ with an opposite direction and a negative label vector.

Based on QG, we are able to compute dimensionalities of crystal structures. The basic formula of dimensionality is[35,36]:

$$\dim(X) = \dim(\{v_1, v_2, \ldots\}) = \operatorname{rank}(V). \tag{1}$$

Here $X$ is a component in which multiple translationally equivalent atoms $\{n_i(0), n_i(v_1), n_i(v_1), \ldots\}$ are connected. $V$ is a matrix composed of the cell offsets between equivalent atoms $\{v_1, v_2, \ldots\}$ and the dimensionality equals to the rank of $V$.

The set of cell offsets is infinite for an extended (1D, 2D and 3D) atomic network, so Eq. (1) is not practical for this situation. To compute the dimensionalities of these infinite systems within finite steps, we should take use of properties of closed chains. A closed chain in a QG represents a path between two equivalent atoms. Suppose these two atoms are $n_{i1}(u_1)$ and $n_{i1}(u_2)$ and the closed chain can be written as:

$$c = n_{i1} \xrightarrow{v_1} n_{i2} \xrightarrow{v_2} n_{i3} \ldots n_{iM} \xrightarrow{v_M} n_{i1}. \tag{2}$$

Because there is an equivalent relation $n_i \xrightarrow{v} n_j \equiv n_j \xrightarrow{-v} n_i$, we can always make all the edges have the same direction. Define the cycle sum of the closed chain $c$: $s(c) = \sum_{k=1}^{M} v_k$. Here $k$ runs over all the edges. It is easy to prove that the cycle sum equals to the cell offset between the pair of equivalent atoms[37]: $s(c) = u_2 - u_1$. For a finite graph, there is a vector space called cycle space composed of all the closed chains. The generating subspace of a cycle space contains finite basic cycles. Therefore, to compute the dimensionality, we only require to calculate cycle sums of basic cycles:

$$\dim(X) = \dim(\{s(c) | c \in F\}) = \operatorname{rank}(S). \tag{3}$$

Here $F$ is the set of basic cycles of the QG and $S$ is a matrix containing all the basic cycle sums.

A crystal structure might contain multiple material components. One material component is corresponding to a connected component in the QG and every material component has independent dimensionality. There are also mixed-dimensional materials containing components with different dimensionalities in databases[36,37].

For detailed discussions about the theory of QG and dimensionality identification, please refer to previous relative works[36–40].

**Molecule identification and searching for molecular crystals**

To search for molecular crystals, a constraint that keeps the bond connectivity of molecules is applied to random generation and evolution processes[1,41]. This constraint can highly reduce the searching space and improve the performance of structure prediction algorithms. For instance, such method is used to explore the complicated pentazolate salts with multiple tens of atoms in the unit cell[42,43]. However, in molecular crystal structure prediction, the molecular configurations are required initially so the constraint may hinder algorithms from finding more stable structures with extended components. For instance, both molecular and chain-like structures are stable in the C-H system[44]. Therefore, using molecular crystal structure prediction alone, we are unable to explore the hydrocarbon phase diagram thoroughly. Furthermore, the constrained searching is not suitable for mixed-dimensional materials like the $P2_1/m$ phase of ammonia[45]. On the other hand, unconstrained crystal structure prediction is more general but it is inefficient for molecular crystal systems. We hope to propose a new evolution scheme which combines the advantages of the constrained and unconstrained structure prediction methods. Here are two problems to be solved. The first one is how to identify the positions and local configurations of molecules in periodic structures. The other is how to treat both isolated and extended structures in a unified framework.

For the first problem, we must consider those bonds across cell boundaries, as shown in Fig. 1(a). If atoms in the same cell are regarded to compose the molecule, it leads to a wrong configuration. A very simple solution to this structure is to consider interatomic distances with periodic boundary condition (PBC). Since $O_2$(-1, 1, -1) (black label) is nearest to $O_1$(0,0,0) (black label) among all the $O_2$ images, they belong to the same molecule. But this approach fails in complex situations such as the structure shown in Fig. 1(b), in which we add two oxygen atoms. It is obvious that the distance between $O_1$(0,0,0) (black label) and $O_4$(0,0,0) (red label) is shorter than that between $O_1$(0,0,0) and $O_4$(-1, 1, -1) (black label). However, $O_1$(0,0,0) is actually connected to $O_4$(-1, 1, -1) (black label) rather than $O_4$(0,0,0) (red label). Actually, a QG-based method described in following is necessary for identifying right molecular structures because the QG has across-boundary information (Fig. 1(c)). Given a 0D connected QG with $N$ vertices, firstly we select an arbitrary vertex marked as $n_0$. Then, for every other vertex $n_M$, we find a path connected to $n_0$: $n_0 \xrightarrow{v_1} ... \xrightarrow{v_M} n_M$. Sum all the edge vectors and we get $\boldsymbol{u}_M = \sum_{j=1}^{M} \boldsymbol{v}_j$. According to the definition of QG, atom $n_M(\boldsymbol{u}_M)$ in cell $\boldsymbol{u}_M$ is connected to atom $n_0(\boldsymbol{0})$ in the original cell. After considering all the vertices, we have a collection $\{n_0(\boldsymbol{0}), n_1(\boldsymbol{u}_1), ..., n_N(\boldsymbol{u}_N)\}$ in which all the atoms belong to the same molecule. For instance, using the quotient graph in Fig. 1(c), we choose $O_1$ as the starting point $n_0$, and get $\boldsymbol{u}_2 = (-1,1,-1)$ for $O_2$. So atom $O_1$ in the original cell and $O_2$ in (-1, 1, -1) cell belong to the same molecule. And then the central position and the local configuration of the molecule are accessible. Such approach can also identify the chain-like molecule in Fig. 1(b) based on the quotient graph in Fig. 1(d).

Note that the selection of path is arbitrary. Suppose there are two different paths connecting two atoms in a molecule and their cell offsets are $\boldsymbol{u}$ and $\boldsymbol{u}'$, respectively. These two paths can compose a closed chain $\boldsymbol{c}$ and cycle sum should be $\boldsymbol{0}$, because the connected component is 0D. So we have $s(\boldsymbol{c}) = \boldsymbol{u} - \boldsymbol{u}' = \boldsymbol{0}$ and $\boldsymbol{u} = \boldsymbol{u}'$. Therefore, the selection of paths does not affect the results.

To address the second issue, the way to treat material components is dependent on their dimensionalities. For a periodic structure, we build QG at first and then identify connected components of the QG. The dimensionality of every component is computed.

If it is 0D, the component is regarded as a whole unit, such as a molecule (Fig. 2(a)). Otherwise, each atom in the component is treated as an isolated part (Fig. 2(b)). For molecular crystals, the scheme maintains the bond connectivity of molecules during heredity and mutation processes just like constrained crystal structure prediction. For extended solids, atoms are independent so the behavior is similar to that of unconstrained searching.

For convenience, we mark this evolution scheme as scheme-1 in this article.

**Decomposing periodic atomic networks**

The scheme-1 described above can automatically detect molecules in crystals, so it is able to accelerate molecular crystal searching. However, it does not affect the searching for extended crystals. Based on scheme-1, we propose another evolution scheme which can also improve searching efficiency for extended systems.

The scheme is based on community detection[46,47]. In general, community detection aims to find a natural way to split networks to communities, and the links inside communities should be much denser than those between communities. In this work, to decompose periodic atomic networks, we take use of Girvan-Newman(GN) algorithm. GN algorithm is top-to-down, which calculates "betweenness" of all the edges and remove the edge with the highest betweenness to generate new communities. This process cannot tell us when to stop removing edges, so in GN algorithm, modularity is used to measure the quality of partitions. Modularity has different forms and parameters, which affect the results of community detection algorithms[48]. But for crystal structure prediction, the problem is not hard to solve. At first, we hope that the communities are 0D so that we can treat them like molecules. In addition, to reduce the searching space, the communities should be as larger as possible. Therefore, GN algorithm should stop when all the communities are 0D in our evolution scheme. The detailed algorithm is shown in Table 1. The input to the algorithm is a material component $G$, which is a connected QG (line 1). We create a set of communities $C$ (line 2). Then the dimensionality of $G$ is computed. If it is 0D (line 3), we add it into $C$ (line 4). If $G$ is an extended component (line 5), we apply GN algorithm to $G$ and consider all the communities (line 6). For every community $c$, we search for its communities and add the results into $C$ (line 7). At the end, the set of all the components is returned (line 9). This algorithm is recursive and the recursion stops when all the components are 0D.

Here we take α-B as an example to demonstrate the processes of our algorithm, as shown in Fig. 3. In the first step (solid arrows), the 3D boron network is split to a 2D $B_{24}$ network and a 0D $B_{12}$ cluster. Then, in the second step (dashed arrows), the 2D network is further decomposed into two 0D $B_{12}$ clusters. Since all the communities are 0D now, the algorithm stops. At the end, the algorithm concludes that $\alpha$-B is composed by three $B_{12}$ icosahedrons. This result is consistent with human's intuition. Ahnert *et al*.[34] have proposed a decomposition method which can also find $B_{12}$ icosahedra in $\alpha$-B. Their method is based on optimizations from random initial partitions while our method is a deterministic algorithm which has fewer parameters.

After decomposition, a crystal structure is regarded to be composed of clusters. This description largely diminishes degrees of freedom of the searching space. For example, previous works generated random structures based on clusters to accelerate predictions of complex structures like allotropes of boron or phosphorus[25,34,41]. Here we applied the decomposition method to the crossover-mutation process in evolutionary structure prediction. The local configurations of clusters are fixed under crossover and mutation operations just like molecular crystals. We mark this evolution scheme as scheme-2. Note that scheme-2 is compatible with scheme-1 for molecular crystals because the method does not decompose 0D material components.

**Results of tests**

To validate the effectiveness of our methods, we select four representative systems for test, as shown in Fig. 4. We mark the standard evolution scheme as the conventional scheme and compare the searching results of conventional scheme, scheme-1 and scheme-2. The searching parameters of all the tests are shown in Table 2.

At first, we compare efficiencies of the schemes in prediction of methane crystal. Under 30 GPa, the most stable phase of methane is $P2_12_12_1$ CH$_4$ [44,49], which is a typical molecular crystal (Fig. 4(a)). Because its symmetry is lower than other high pressure phases, it is relatively harder to find it.

In this test, we performed 10 independent global optimizations under 30 GPa with conventional, scheme-1 and scheme-2 evolution schemes, respectively. The maximum number of generations is 30 and 30 structures are relaxed in every generation. The searching results of both conventional scheme and scheme-1 are shown in Table 3. For conventional scheme, only 7 tests (70%) find the $P2_12_12_1$ phase, while our scheme-1 and scheme-2 get success in all the tests. The results show that the molecule identification method can help evolution operators to generate more reasonable candidate structures for molecular crystals, which improves the success rate of evolutionary searching. The new schemes also reduce the number of local optimizations before finding the global minimum. The improvement induced by scheme-1 is not very large (10%) while the performance of scheme-2 is much better (reduce the number of required structures by about 35% compared with the conventional method). Since there are no constraints in the structure generation process, some relaxed configurations are extended networks. For these structures, scheme-2 can decompose them more reasonably than scheme-1.

Molecule identification can also accelerate searching for mix-dimensional structures, which is beyond the scope of molecular crystal structure prediction. The $P2_1/m$ phase of ammonia[45,50] is chosen to be an example for test. It is composed of 0D $NH_4^+$ ions and 1D $NH_2^-$ chains (Fig. 4(b)). The stable pressure range of the $P2_1/m$ phase is from 343 to 450 GPa based on accurate DFT calculations[50]. In this work, ammonia structures are relaxed using DFT with low accuracy parameters. It leads to a deviation from accurate results and $P2_1/m$ phase is stable under 300 GPa. Here we have also run 10 independent tests for ammonia at 300 GPa with all the evolution schemes. The maximum number of generations and population size are the same to those of tests for methane. The results are shown in Table 3. For this structure containing both isolated and extended components, conventional scheme is able to find the global minimum in all the tests. But its efficiency is obviously lower than scheme-1 and scheme-2. Scheme-1 and scheme-2 reduce the number of required relaxations by about 30% and 45%, respectively. They show significant advantages against the standard evolution scheme. For systems containing extended nets like this, community detection is also useful, so the number of required structures for scheme-2 is less than that for scheme-1 by about 23%.

The tests above are designed for materials containing isolated components, in which molecule identification can be used for improving structure searching. In the following, we will only compare the efficiency of conventional and scheme-2 evolutions schemes in extended test systems.

$MgAl_2O_4$ is a popular system to validate structure prediction algorithms[32,51] (Fig. 4 (c)). Because the structures of this system can be relaxed based on a fast classical force field, extensive tests (50 times) under 100 GPa are performed with conventional scheme and scheme-2, respectively. The maximum number of generations is 60 and 50 structures are relaxed in every generation. For conventional scheme, 512 structures are required to find *Pnma* phase while 378 structures are required for scheme-2. The number of required structures decreases by about 26% with scheme-2. It indicates that community detection can reduce the phase space of solids and improve the efficiency of searching. This result is very closed to the best results reported for this system[32] (368 structures). It is very interesting that this record is created also by a graph-theory based method. Although in the previous work they applied graph theory to the structure generation process while our method aims at the crossover and mutation stage. Therefore, there are no conflict between them, and both cases demonstrate that the applications of graph theory are very helpful for crystal structure prediction. It is expected that the combination of these graph-theory based methods may further improve searching efficiency.

Recently, many efforts have been devoted to searching for complex boron structures using machine learning techniques[22–24]. Here we also choose $\gamma$-B[17] as a more complicated example to test our method further (Fig. 4(d)). To improve the success rate, random structures are generated with a fixed space group *Pnnm* for both conventional and scheme-2 evolution schemes. We carried out 20 independent tests with a maximum number of generations of 60 and a population size of 50. A machine-learning potential for boron[24] is used for structure relaxations. The results are shown in Table 3. It is not surprising that scheme-2 not only increases the success rate but also dramatically reduces the number of structures required for discovering $\gamma$-B by more than 40%. Compared with $MgAl_2O_4$, boron allotropes have more distinct community structures, so the decomposition algorithm can identify larger clusters and make more constraints to the structural space. Therefore, the effects of scheme-2 for boron is much larger than those for $MgAl_2O_4$. Ahnert *et al*. performed random structure searching for boron based on community detection[34]. Their method decomposes networks to clusters and builds random structures using these clusters for prediction while our on-the-fly method is applied to the evolution process. Our results indicate that graph theory is a very useful tool for prediction of complex crystals. In addition, we have applied the new evolution scheme to find novel mixed coordination silica and silicon superoxides.[52]

In summary, we have developed two new evolution schemes aided by graph theory for crystal structure predictions. We have discussed how to identify molecules in crystals using quotient graph and decompose extended structures reasonably in detail. These new schemes, based on dimensionality identification and community detection, are designed to find molecules and clusters in periodic structures automatically. During the crossover-mutation process, the local configurations of these molecules or clusters are kept so as to reduce the phase space implicitly. With representative examples for test, we have demonstrated their capabilities of improving success rates and the searching efficiency for both isolated and extended systems.

**METHODS**

Local relaxations of methane and ammonia are performed using the Vienna ab initio simulation (VASP) code[53] with projector augmented-wave potentials and the Perdew-Burke-Ernzerhof exchange-correlation functional (PBE)[54]. The accuracy of DFT calculations is not very high in this work so that the time cost of global optimizations is affordable. For $MgAl_2O_4$, we use Lewis–Catlow potentials[55] implemented in GULP[56] to perform local optimizations. For $\gamma$-B, we conduct a Gaussian approximation potential trained for boron[24,57] to relax structures. The global prediction is performed by our in-house code. For every test system, we only make the minimal modification on parameters for different evolution schemes and most of the parameters are the same.

**ACKNOWLEDGMENTS**

The authors thank Chris Pickard for fruitful discussions. J.S. gratefully acknowledges financial support from the National Key R&D Program of China (grant nos. 2016YFA0300404), the National Natural Science Foundation of China (grant nos. 11974162 and 11834006), and the Fundamental Research Funds for the Central Universities. The calculations were carried out using supercomputers at the High Performance Computing Center of Collaborative Innovation Center of Advanced Microstructures, the high-performance supercomputing center of Nanjing University, and 'Tianhe-2' at NSCC-Guangzhou.

# REFERENCES

1. Pickard, C. J. & Needs, R. J. *Ab initio* random structure searching. *J. Phys.: Condens. Matter* **23**, 053201 (2011).

2. Glass, C. W., Oganov, A. R. & Hansen, N. USPEX—Evolutionary crystal structure prediction. *Computer Physics Communications* **175**, 713–720 (2006).

3. Lonie, D. C. & Zurek, E. XtalOpt: An open-source evolutionary algorithm for crystal structure prediction. *Computer Physics Communications* **182**, 372–387 (2011).

4. Yao, Y., Klug, D. D., Sun, J. & Martoňák, R. Structural Prediction and Phase Transformation Mechanisms in Calcium at High Pressure. *Physical Review Letters* **103**, (2009).

5. d'Avezac, M. & Zunger, A. Identifying the minimum-energy atomic configuration on a lattice: Lamarckian twist on Darwinian evolution. *Physical Review B* **78**, (2008).

6. Tipton, W. W. & Hennig, R. G. A grand canonical genetic algorithm for the prediction of multi-component phase diagrams and testing of empirical potentials. *J. Phys.: Condens. Matter* **25**, 495401 (2013).

7. Wang, Y., Lv, J., Zhu, L. & Ma, Y. Crystal structure prediction via particle-swarm optimization. *Physical Review B* **82**, (2010).

8. Wales, D. J. & Doye, J. P. K. Global Optimization by Basin-Hopping and the Lowest Energy Structures of Lennard-Jones Clusters Containing up to 110 Atoms. *J. Phys. Chem. A* **101**, 5111–5116 (1997).

9. Goedecker, S. Minima hopping: An efficient search method for the global minimum of the potential energy surface of complex molecular systems. *The Journal of Chemical Physics* **120**, 9911–9917 (2004).

10. Schön, J. C. & Jansen, M. First Step Towards Planning of Syntheses in Solid-State Chemistry: Determination of Promising Structure Candidates by Global Optimization. *Angewandte Chemie International Edition in English* **35**, 1286–1304 (1996).

11. Martoňák, R., Laio, A. & Parrinello, M. Predicting Crystal Structures: The Parrinello-Rahman Method Revisited. *Phys. Rev. Lett.* **90**, 075503 (2003).

12. Xia, K. *et al.* A novel superhard tungsten nitride predicted by machine-learning accelerated crystal structure search. *Sci. Bull.* **63**, 817–824 (2018).

13. Jørgensen, M. S., Larsen, U. F., Jacobsen, K. W. & Hammer, B. Exploration versus Exploitation in Global Atomistic Structure Optimization. *J. Phys. Chem. A* **122**, 1504–1509 (2018).

14. Yamashita, T. *et al.* Crystal structure prediction accelerated by Bayesian optimization. *Phys. Rev. Materials* **2**, 013803 (2018).

15. Zhang, L., Wang, Y., Lv, J. & Ma, Y. Materials discovery at high pressures. *Nature Reviews Materials* **2**, 17005 (2017).

16. Oganov, A. R., Pickard, C. J., Zhu, Q. & Needs, R. J. Structure prediction drives materials discovery. *Nature Reviews Materials* **4**, 331 (2019).

17. Oganov, A. R. *et al.* Ionic high-pressure form of elemental boron. *Nature* **457**, 863–867 (2009).

18. Zhang, W. *et al.* Unexpected Stable Stoichiometries of Sodium Chlorides. *Science* **342**, 1502–1505 (2013).

19. Wang, X. *et al.* Pressure-induced structural and electronic transitions in bismuth iodide. *Phys. Rev. B* **98**, 174112 (2018).

20. Salke, N. P. *et al.* Tungsten hexanitride with single-bonded armchair-like hexazine structure at high pressure. *Phys. Rev. Lett. (accepted)*.

21. Jacobsen, T. L., Jørgensen, M. S. & Hammer, B. On-the-Fly Machine Learning of Atomic Potential in Density Functional Theory Structure Optimization. *Phys. Rev. Lett.* **120**, 026102 (2018).

22. Tong, Q., Xue, L., Lv, J., Wang, Y. & Ma, Y. Accelerating CALYPSO Structure Prediction by Data-driven Learning of Potential Energy Surface. *Faraday Discussions* (2018) doi:10.1039/C8FD00055G.

23. Podryabinkin, E. V., Tikhonov, E. V., Shapeev, A. V. & Oganov, A. R. Accelerating crystal structure prediction by machine-learning interatomic potentials with active learning. *Phys. Rev. B* **99**, 064114 (2019).

24. Deringer, V. L., Pickard, C. J. & Csányi, G. Data-Driven Learning of Total and Local Energies in Elemental Boron. *Phys. Rev. Lett.* **120**, 156001 (2018).

25. Deringer, V. L., Proserpio, D. M., Csanyi, G. & Pickard, C. J. Data-driven learning and prediction of inorganic crystal structures. *Faraday Discussions* (2018) doi:10.1039/C8FD00034D.

26. Bisbo, M. K. & Hammer, B. Efficient Global Structure Optimization with a Machine-Learned Surrogate Model. *Phys. Rev. Lett.* **124**, 086102 (2020).

27. Chung, S. J., Hahn, Th. & Klee, W. E. Nomenclature and generation of three-periodic nets: the vector method. *Acta Cryst. A* **40**, 42–50 (1984).

28. O'Keeffe, M., Eddaoudi, M., Li, H., Reineke, T. & Yaghi, O. M. Frameworks for Extended Solids: Geometrical Design Principles. *Journal of Solid State Chemistry* **152**, 3–20 (2000).

29. Blatov, V. A., Shevchenko, A. P. & Proserpio, D. M. Applied Topological Analysis of Crystal Structures with the Program Package ToposPro. *Cryst. Growth Des.* **14**, 3576–3586 (2014).

30. Shi, X., He, C., Pickard, C. J., Tang, C. & Zhong, J. Stochastic generation of complex crystal structures combining group and graph theory with application to carbon. *Phys. Rev. B* **97**, 014104 (2018).

31. He, C. *et al.* Complex Low Energy Tetrahedral Polymorphs of Group IV Elements from First Principles. *Phys. Rev. Lett.* **121**, 175701 (2018).

32. Bushlanov, P. V., Blatov, V. A. & Oganov, A. R. Topology-based crystal structure generator. *Computer Physics Communications* **236**, 1–7 (2019).

33. Alexandrov, E. V., Shevchenko, A. P., Asiri, A. A. & Blatov, V. A. New knowledge and tools for crystal design: local coordination versus overall network topology and much more. *CrystEngComm* **17**, 2913–2924 (2015).

34. Ahnert, S. E., Grant, W. P. & Pickard, C. J. Revealing and exploiting hierarchical material structure through complex atomic networks. *npj Comput. Mater.* **3**, 35 (2017).

35. Mounet, N. *et al.* Two-dimensional materials from high-throuput computational exfoliation of experimentally known compounds. *Nat. Nanotechnol.* **13**, 246–252 (2018).

36. Larsen, P. M., Pandey, M., Strange, M. & Jacobsen, K. W. Definition of a scoring parameter to identify low-dimensional materials components. *Phys. Rev. Materials* **3**, 034003 (2019).

37. Gao, H., Wang, J., Guo, Z. & Sun, J. Determining dimensionalities and multiplicities of crystal nets. *npj Computational Materials* **6**, 1–9 (2020).

38. Blatov, V. A., Carlucci, L., Ciani, G. & Proserpio, D. M. Interpenetrating metal–organic and inorganic 3D networks: a computer-aided systematic investigation. Part I. Analysis of the Cambridge structural database. *CrystEngComm* **6**, 377–395 (2004).

39. Thimm, G. A graph theoretical approach to the analysis, comparison, and enumeration of crystal structures. (2008).

40. Eon, J.-G. Topological features in crystal structures: a quotient graph assisted analysis of underlying nets and their embeddings. *Acta Cryst. A* **72**, 268–293 (2016).

41. Zhu, Q., Oganov, A. R., Glass, C. W. & Stokes, H. T. Constrained evolutionary algorithm for structure prediction of molecular crystals: methodology and applications. *Acta Cryst. A* **68**, 215–226 (2012).

42. Xia, K. *et al.* Pressure-Stabilized High-Energy-Density Alkaline-Earth-Metal Pentazolate Salts. *J. Phys. Chem. C* **123**, 10205–10211 (2019).

43. Xia, K. *et al.* Predictions on High-Power Trivalent Metal Pentazolate Salts. *J. Phys. Chem. Lett.* **10**, 6166–6173 (2019).

44. Conway, L. J. & Hermann, A. High Pressure Hydrocarbons Revisited: From van der Waals Compounds to Diamond. *Geosciences* **9**, 227 (2019).

45. Pickard, C. J. & Needs, R. J. Highly compressed ammonia forms an ionic crystal. *Nature Mater* **7**, 775–779 (2008).

46. Girvan, M. & Newman, M. E. J. Community structure in social and biological networks. *PNAS* **99**, 7821–7826 (2002).

47. Newman, M. E. J. & Girvan, M. Finding and evaluating community structure in networks. *Physical Review E* **69**, (2004).

48. Fortunato, S. Community detection in graphs. *Physics Reports* **486**, 75–174 (2010).

49. Gao, G. *et al.* Dissociation of methane under high pressure. *J. Chem. Phys.* **133**, 144508 (2010).

50. Griffiths, G. I. G., Needs, R. J. & Pickard, C. J. High-pressure ionic and molecular phases of ammonia within density functional theory. *Phys. Rev. B* **86**, (2012).

51. Lyakhov, A. O., Oganov, A. R., Stokes, H. T. & Zhu, Q. New developments in evolutionary structure prediction algorithm USPEX. *Computer Physics Communications* **184**, 1172–1182 (2013).

52. Liu, C. *et al.* Mixed Coordination Silica at Megabar Pressure. *Phys. Rev. Lett.* **126**, 035701 (2021).

53. Kresse, G. & Furthmüller, J. Efficient iterative schemes for *ab initio* total-energy calculations using a plane-wave basis set. *Phys. Rev. B* **54**, 11169–11186 (1996).

54. Perdew, J. P., Burke, K. & Ernzerhof, M. Generalized Gradient Approximation Made Simple. *Phys. Rev. Lett.* **77**, 3865–3868 (1996).

55. Lewis, G. V. & Catlow, C. R. A. Potential models for ionic oxides. *J. Phys. C: Solid State Phys.* **18**, 1149–1161 (1985).

56. Gale, J. D. & Rohl, A. L. The General Utility Lattice Program (GULP). *Molecular Simulation* **29**, 291–341 (2003).

57. Bartók, A. P., Payne, M. C., Kondor, R. & Csányi, G. Gaussian Approximation Potentials: The Accuracy of Quantum Mechanics, without the Electrons. *Physical Review Letters* **104**, 136403 (2010).

58. Jain, A. *et al.* Commentary: The Materials Project: A materials genome approach to accelerating materials innovation. *APL Materials* **1**, 011002 (2013).

**Figures and Tables**

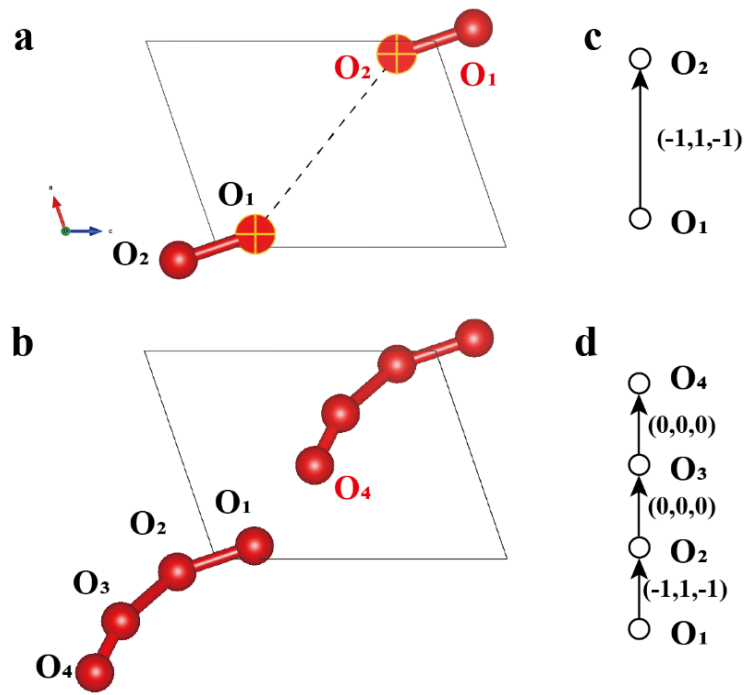

Fig. 1 (a) An oxygen structure in Material Project[58] (ID: mp-1009490). The marked atoms are in the same cell and the dashed line is the wrong connection. (b) Add two atoms into the structure in (a). The atoms in same molecule have labels with same color. (c) The QG of the structure in (a). (d) The QG of the structure in (b).

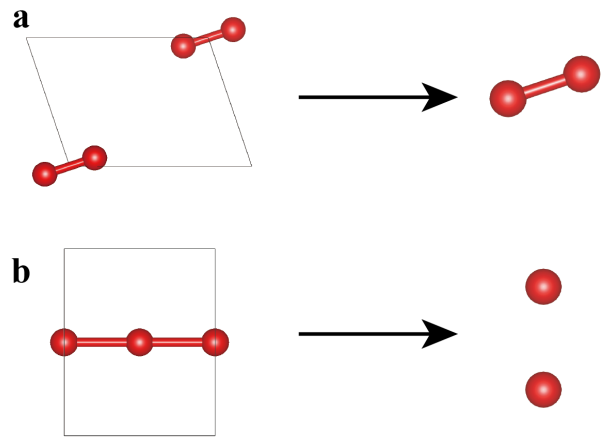

Fig. 2 (a) For a molecular crystal, the molecule is extracted. (b) For an extended crystal, all the atoms are extracted independently.

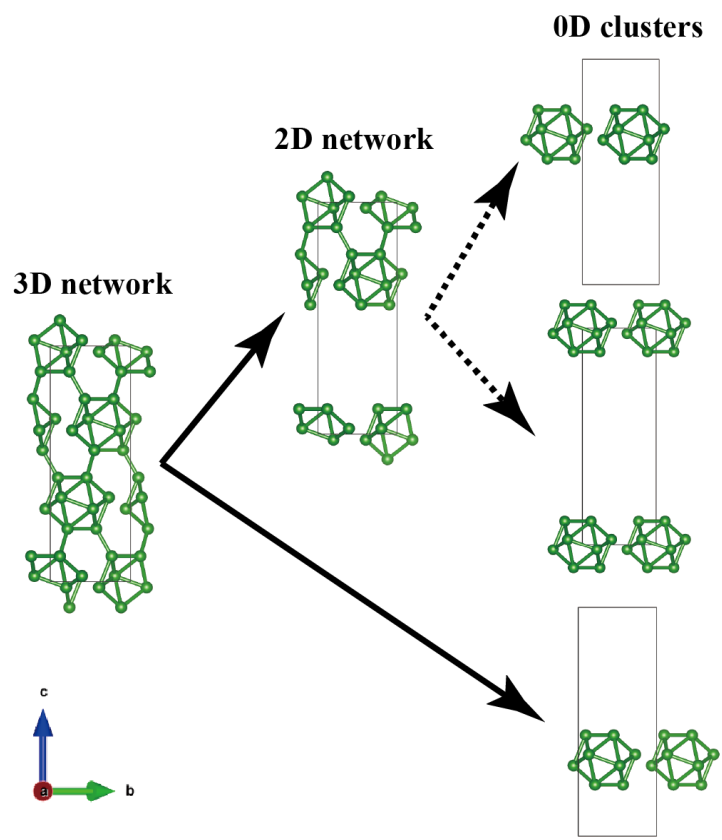

Fig. 3 The decomposition processes of α-B using our algorithm. Solid and dashed arrows represent the first and second decomposition steps, respectively.

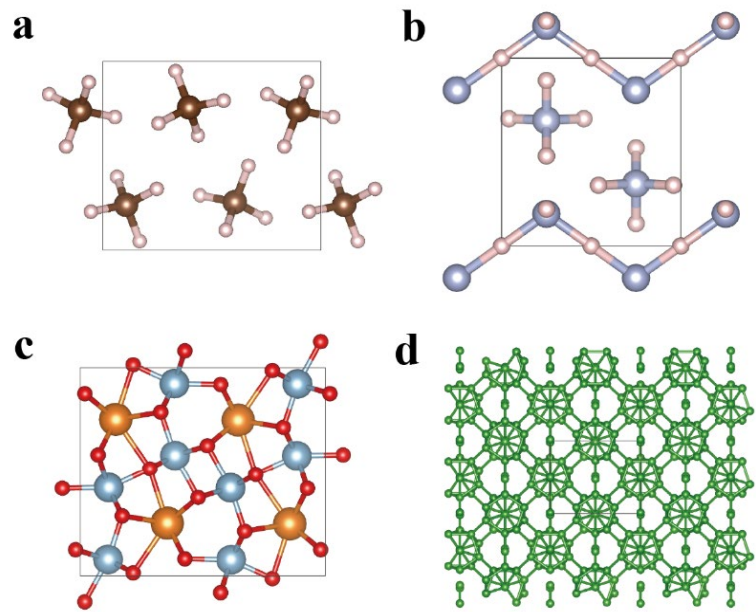

Fig. 4 Examples for test in this work. (a) $P2_12_12_1$ CH$_4$ (20 atoms in the unit cell) (b) $P2_1/m$ NH$_3$ (16 atoms in the unit cell) (c) $Pnma$ MgAl$_2$O$_4$ (28 atoms in the unit cell) (d) $Pnnm$ $\gamma$-B (28 atoms in the unit cell).

**Table 1. Pseudocode for finding largest communities from a connected component.**

| | |
|---|---|
| 1: | **procedure** FINDCOMMUNITIES(*G*) |
| 2: |   *C* := ∅ |
| 3: |   **if** Dim(*G*)= 0 **then** |
| 4: |     *C* := *C* ∪ {*G*} |
| 5: |   **else** |
| 6: |     **for** *c* ∈ Girvan-Newman(*G*) **do** |
| 7: |       *C* := *C* ∪ FINDCOMMUNITIES(*c*) |
| 8: |   **end if** |
| 9: |   **return** *C* |
| 10: | **end procedure** |

**Table 2. Parameters of searching. $N_{at}$: number of atoms; $N_t$: number of independent tests; $N_g$: maximum number of generations; $N_p$: population size.**

| System | $N_{at}$ | $N_t$ | $N_g$ | $N_p$ | Pressure (GPa) |
|---|---|---|---|---|---|
| $CH_4$ | 20 | 10 | 30 | 30 | 30 |
| $NH_3$ | 16 | 10 | 30 | 30 | 300 |
| $MgAl_2O_4$ | 28 | 50 | 60 | 50 | 100 |
| B (fixed symmetry) | 28 | 20 | 60 | 50 | 100 |

**Table 3. Results of tests. $N_{avg}$: average number of structures until global minimum is found.**

| System | Evolution scheme | Success rate(%) | $N_{avg}$ |
|---|---|---|---|
| $CH_4$ | conventional | 70 | 270 |
| | scheme-1 | 100 | 243 |
| | scheme-2 | 100 | 174 |
| $NH_3$ | conventional | 100 | 342 |
| | scheme-1 | 100 | 243 |
| | scheme-2 | 100 | 186 |
| $MgAl_2O_4$ | conventional | 100 | 512 |
| | scheme-2 | 100 | 378 |
| B (fixed symmetry) | conventional | 75 | 1300 |
| | scheme-2 | 100 | 775 |